\documentclass[12pt,epsf]{article}
\usepackage{graphicx}
\usepackage{epsf}
\begin{document}
\newcommand{\sheptitle}
{ Deviation from Tri-bimaximal Mixings in Two Types of Inverted Hierarchical Neutrino Mass Models }
\newcommand{\shepauthor}
{N.Nimai Singh$^{a,}$\footnote{Regular Associate , The Abdus Salam ICTP, Trieste, Italy,
\\ e-mail: nimai03@yahoo.com}, Monisa Rajkhowa$^{a, b}$
 and Abhijit Borah$^{c}$}
\newcommand{\shepaddress}
{$^{a}$ Department of Physics, Gauhati University, Guwahati-781014, India\\
$^{b}$ Department of Physics, Science College, Jorhat-781014, Assam, India\\
$^{c}$ Department of Physics, Fazl Ali College, Mokokchung-798601, Nagaland, India}

\newcommand{\shepabstract}
{An attempt is made  to explore the possibility  for  deviations of 
solar mixing angle ($\theta_{12}$)
from  tri-bimaximal mixings, without sacrificing the predictions of maximal 
atmospheric mixing angle ($\theta=\pi/4$) and zero reactor angle ($\theta_{13}=0$).
We find that the above conjecture  can be automatically realised in the  inverted 
hierarchical neutrino mass model
 having 2-3 symmetry, in the basis where  charged lepton mass matrix is diagonal.
For the observed ranges of $\bigtriangleup m_{21}^2$ and $\bigtriangleup m_{23}^2$,
we calculate the predictions on  $\tan^2\theta_{12}=0.5, 0.45, 0.35$ for  different 
input values of the parameters in the neutrino mass matrix. We also observe  a 
 possible crossing over
 from one type of inverted hierarchical model having same CP parity (Type-IHA) to other 
type having opposite CP parity (Type-IHB). Such neutrino mass matrices can be  obtained  
from the canonical seesaw formula using diagonal form of Dirac neutrino mass matrix and 
non-diagonal texture of right-handed Majorana mass matrix, and  may  have important implications 
in model building using discrete as well as non-abelian symmetry groups.     \\ \\
Keywords: inverted hierarchical, tri-bimaximal mixings, solar mixing angle,neutrino mass matrix.\\ \\
PACS numbers:14.60.Pq,12.15.Ff,13.15.+g,13.40.Em}
\begin{titlepage}
\begin{flushright}
\end{flushright}
\begin{center}
{\large{\bf\sheptitle}} 
\bigskip\\
\shepauthor
\\
\mbox{}
\\
{\it\shepaddress}
\\
\vspace{.5in}
{\bf Abstract}\bigskip\end{center}\setcounter{page}{0}
\shepabstract
\end{titlepage}
\section{Introduction}
Current observational data[1] on neutrino oscillations indicates a clear departure
 from the tri-bimaximal mixings or Harrison-Perkins-Scott (HPS) mixing pattern[2]. 
The most recent SNO experimental determination[3] of solar angle gives 
$\tan^2\theta_{12}=0.45^{+0.09}_{-0.08}$ compared with  $\tan^2\theta_{12}=0.50$ in 
HPS scheme. There is no  strong claim for a substantial departure from the maximal 
atmospheric mixing  
 ($\tan^2\theta_{23}=1$),  and also from the zero reactor angle ($\sin\theta_{13}=0$).
Only upper bound for  $\sin\theta_{13}$ is known at the moment and 
future measurements may possibly give  a very small value so small that we can
 approximate it by zero[4]. This does  not yet contradict with the  non-observation of 
Dirac CP phase angle. There are discussions[5] on the experimental requirements for mass
 hierarchy measurements for $\theta_{13}=0$.

Conditions of maximal atmospheric mixing $(\theta_{23}=\pi/4)$
 and exact zero of reactor angle $(\theta_{13}=0)$, are in fact 
necessary and sufficient conditions[4-9] for the leptonic mixing matrix obtained 
from the diagonalization of the left-handed Majorana neutrino mass
 matrix having a 2-3 symmetry. Here the  2-3 symmetry simply implies an invariance 
under the simultaneous permutation of the second and third rows as well as 
the second and third columns[4]. In the basis where charged lepton mass matrix 
is diagonal, a 2-3 symmetry can be generally realised in three degenerate
 models, 
and two inverted hierarchical models of neutrino mass patterns[10].
 However, normal hierarchical model 
generally does not possess a  2-3 symmetry and hence it does not predict maximal atmospheric 
mixing as well as exact zero reactor angle[11]. The degenerate models on the other hand 
give either very small ($\tan^2\theta_{12}\approx 0.27$) or maximal solar mixings[10]. 
 Further constraints such as zero determinant[12,13,4] 
 or zero trace[14] of the neutrino mass matrix, which lead to other interesting 
properties, are not considered in the present analysis. 
This freedom  allows us to consider  larger value of   non-zero mass eigenvalue $m_3$
 within the framework of  inverted 
hierarchical model.

A general form of inverted hierarchical  mass matrix $m_{LL}$ having 2-3 symmetry,
 can be written as[4],    
 \begin{equation}
m_{LL}= 
\left(\begin{array}{ccc}
m_{11} & m_{12} & m_{12} \\
m_{12} & m_{22} & m_{23} \\
m_{12} & m_{23} & m_{22}
\end{array}\right)
\end{equation}
which is  diagonalised by the relation $ m_{LL}=UDU^{\dag}$ where $U$ is given by  
\begin{equation}
U= 
\left(\begin{array}{ccc}
c_{12} & -s_{12} & 0 \\
\frac{s_{12}}{\sqrt{2}} & \frac{c_{12}}{\sqrt{2}} & -\frac{1}{\sqrt{2}}\\
\frac{s_{12}}{\sqrt{2}} & \frac{c_{12}}{\sqrt{2}} & \frac{1}{\sqrt{2}}
\end{array}\right).
\end{equation}
Here we have  $c_{12}=\cos\theta_{12}$ and $s_{12}=\sin\theta_{12}$; 
and $\theta_{23}=\pi/4$ and $\theta_{13}=0$.
In the basis where charged lepton mass matrix  is  diagonal, $U$ in eq.(2) 
is identified as the MNS mixing 
matrix $U_{MNS}$[15] where the solar mixing angle $\theta_{12}$ is arbitrary, atmospheric 
mixing angle $\theta_{23}$ is maximal and reactor angle $\theta_{13}$ is exactly zero.
For bimaximal mixings, we choose  $c_{12}=1/\sqrt 2$ and $s_{12}=1/\sqrt 2$, leading to 
$\tan^2\theta_{12}=1.0$, whereas for tri-bimaximal mixings[2], we have  
$c_{12}=\sqrt {\frac{2}{3}}$ and $s_{12}=1/\sqrt 3$, leading to 
$\tan^2\theta_{12}=0.5$.

We are now interested in the present analysis to investigate the condition for 
fixing the arbritary  solar mixing angle 
to its tri-bimaximal value[2] and also for a possible  gradual departure from this 
tri-bimaximal  value, without sacrificing maximal atmospheric mixing  and exact 
zero value of reactor angle. We  calculate here  three neutrino mass eigenvalues 
 consistent with the present observational  data[1].
  
 We  confine our analysis in  inverted hierarchical neutrino 
mass model having a 2-3 symmetry in eq.(1). We organise the paper as follows.
 In section 2 we present an analytic solution for the inverted hierarchical model
 and find out condition 
for lowering the solar mixing angle. This is followed by  a crossing over from  
 Type-IHA to Type-IHB inverted hierarchical models.
Numerical results  are presented in  section 3. We conclude in section 4 with a summary and discussions. 

\section{Analysis of Inverted hierarchical model}
We have in general two  types[16,10,11] of inverted hierarchical models based on 
the relative sign of the first 
two mass  eigenvalues $m_1$ and $ m_2$ : Type-IHA for same CP parity $(m_1, m_2, m_3)$ and Type-IHB
 for opposite CP parity  $(m_1, - m_2, m_3)$. Type-IHB is considered to be more  stable under
 radiative corrections in MSSM [17-19],  whereas Type-IHA is found to be more stable under the presence 
of left-handed 
Higgs triplet term in type-II seesaw mechanism[20]. For our present  analysis we will not 
address the issue 
of stability of neutrino mass model.  Instead, we explore the properties of these two types 
of inverted hierarchical mass matrices and their interconnection. 

\subsection{Inverted hierarchy of Type-IHA}
 We start with a specific form[10] of Type-IHA mass matrix having 2-3 symmetry (1) as, 
  \begin{equation}
M_{IHA}= 
\left(\begin{array}{ccc}
1-2\epsilon & -\epsilon & -\epsilon \\
-\epsilon & \frac{1}{2} & \frac{1}{2}-\eta \\
-\epsilon & \frac{1}{2}-\eta & \frac{1}{2}
\end{array}\right)m_0,
\end{equation}
where the symmetry breaking parameters are $\epsilon,\eta << 1$ 
and $m_{0}=0.05$eV as imput value[10]. Such left-handed Majorana mass matrix 
can be realised in the canonical seesaw formula using a generalised diagonal form of 
Dirac mass matrix $m_{LR}$ and non-diagonal form of right-handed Majorana mass matrix $M_{RR}$
 given by[10] 
\begin{equation}
m_{LR}= 
\left(\begin{array}{ccc}
\lambda^m & 0  & 0 \\
0 & \lambda^n & 0 \\
0 & 0 & 1
\end{array}\right)m_f,
\end{equation}
 and 
\begin{equation}
M_{RR}= 
\left(\begin{array}{ccc}
(1+2\epsilon)\lambda^{2m} & \epsilon\lambda^{m+n} & \epsilon\lambda^{m} \\
\epsilon\lambda^{m+n} & \frac{1}{2\eta}\lambda^{2n} &(1-\frac{1}{2\eta})\lambda^n \\
\epsilon\lambda^{m} & (1-\frac{1}{2\eta})\lambda^n & \frac{1}{2\eta}
\end{array}\right)v_0,
\end{equation}
where $m_{0}=m_f^2/v_{0}$. The pair of $(m,n)$ can take different sets of values which can represent
 mass matrices belonging to up-quark, down-quark, charged-lepton or any diagonal form of Dirac neutrino. 

$M_{IHA}$ in eq.(3) can be   reduced to  the zeroth order texture[16] when  $\epsilon=\eta=0$,  
\begin{equation}
M^0_{IHA}= 
\left(\begin{array}{ccc}
1 & 0  & 0 \\
0 & \frac{1}{2} & \frac{1}{2} \\
0 & \frac{1}{2} & \frac{1}{2}
\end{array}\right)m_0,
\end{equation}
This has a degeneracy in the first two  mass eigenvalues, $(1, 1, 0)m_0$, and 
such degeneracy makes the solar mixing angle $\theta_{12}$  arbitrary, and may have infinite  values
lying  between $0$ and $\frac{\pi}{4}$.
Once the degeneracy is removed as in eq.(3), the solar angle is then fixed at a particular value.
 Such freedom 
in fixing the solar angle does not destroy 2-3 symmetry of the mass matrix, and it 
depends absolutely on the  choice of input values of $\eta$ and $\epsilon$,
without disturbing the predictions on  atmospheric angle $\theta_{23}=\frac{\pi}{4}$ and 
reactor angle $\theta_{13}=0$. Diagonalizing (3) we obtain  the three  mass eigenvalues, 
\begin{equation}
m_1=(2 - 2\epsilon - \eta - y)\frac{m_0}{2},\ \ \
 m_2=(2 - 2\epsilon - \eta + y)\frac{m_0}{2},\ \ \
 m_3= \eta m_0
\end{equation}
where 
\begin{equation}
y^2= 12 \epsilon^2 -4 \epsilon \eta + \eta^2.
\end{equation}
The solar mixing is now fixed by the relation[7],
\begin{equation}
\tan 2\theta_{12}= \frac{2\sqrt{2}}{2-\frac{\eta}{\epsilon}}
\end{equation}
As long as $(2-\frac{\eta}{\epsilon})\leq 1$, we have tri-bimaximal (or HPS) mixing  and its 
possible deviation to lower values. For tri-bimaximal mixing, we find
 two sulutions of eq.(7), for  $\eta/\epsilon$ at  $1$ and $3$ respectively. 
Similarly, for deviation from tri-biimaximal solar mixing to lower values, we have the
 corresponding  values of this 
ratios as  $\eta/\epsilon <1$ and  $\eta/\epsilon > 3$. Once the solar angle is fixed, then the range 
of $\eta$ or $\epsilon$ can be solved through a search programme. Table-1 represents a 
summary of our results.

\subsection{Crossing over from Type-IHA to Type-IHB}
 One particular observation for the  solution of eq.(3) in  the range  $\epsilon > 0.5$, is the crossing 
over from inverted hierarchical model  with same CP parity (Type-IHA) to inverted hierarchical model 
 with opposite 
CP parity (Type-IHB). To demonstrate this interconnection, we start with the most general form 
of mass matrix[8] belonging to  Type-IHB,  
 \begin{equation}
M_{IHB}= 
\left(\begin{array}{ccc}
\delta_1 & 1 & 1 \\
1 & \delta_{2} & \delta_{3} \\
1 & \delta_{3} & \delta_{2}
\end{array}\right) m'_{0}.
\end{equation}
The zeroth order mass matrix of eq.(10) has the form[16] 
 \begin{equation}
 M^0_{IHB}=
\left(\begin{array}{ccc}
0  & 1 & 1 \\
1 & 0  & 0  \\
1 & 0  & 0
\end{array}\right)m'_{0},
\end{equation}
which has non-degenerate eigenvalues $(1, -1, 0)m'_{0}$ compared to that of 
 eq.(6). 
In addition, the solar mixing 
is now fixed at  the maximal value ($\theta_{12}=\pi/4$)  unlike eq.(4) where it can have any arbitrary   value.
The diagonalization of (10) leads to the following mass eigenvalues, 
\begin{equation}
m_1=(\delta_1 + \delta_2 + \delta_3 + x)\frac{m'_0}{2},\ \ \
m_2=(\delta_1 + \delta_2 + \delta_3 - x)\frac{m'_0}{2},\ \ \
 m_3=(\delta_2 - \delta_3)m'_0
\end{equation}
where 
\begin{equation}
x^2=8+(\delta_1^2+\delta_2^2+\delta_3^2)-2\delta_1\delta_2-2\delta_1\delta_3+2\delta_2\delta_3;
\end{equation}
and the  solar mixing, 
\begin{equation}
\tan 2\theta_{12}= \frac{2\sqrt{2}}{(\delta_1-\delta_2-\delta_3)}.
\end{equation}
Type-IHB in (10) generally predicts nearly maximal solar mixing and it has been  considered 
almost impossible to tone down the solar angle without corrections from charged 
lepton mass matrix[21]. Here we wish to show that this notion is not completely true  and  
can be avoided through the following transformation, 
\begin{equation}
\delta_1=2(1-\frac{1}{2\epsilon}), \ \ \ 
\delta_2=-\frac{1}{2\epsilon}, \ \ \
\delta_3=(\frac{\eta}{\epsilon}-\frac{1}{2\epsilon}), \ \ \
m'_{0}=m_{0}(-\epsilon)
\end{equation}
where the three parameters $\delta_1, \delta_2,  \delta_3$ are now replaced by only 
two parameters $\eta,  \epsilon$.
From eq.(15) it is easy to see that  inverted 
hierarchical model (Type-IHB)  in eq.(10) is now equivalent to Type-IHA in eq.(3), and 
their expressions for   eigenvalues and
 solar angle  in eqs.(7)-(9) and eqs.(12) - (14) respectively, are also equivalent.
 As emphasised earlier,
such cross-over will take place only when the value of  $\epsilon > 0.5$ and this is shown in Table-1.   
\section{Numerical calculations and results}
We follow two important steps for carrying out numerical estimations.
 In the first step, we choose a specific value of solar mixing 
$\tan^2\theta_{12}$ via eq.(9) or (14), and then solve for possible values of the ratio $\eta/\epsilon$.
In the second step we take up  a particular value of this ratio $\eta/\epsilon$, 
and find out the ranges 
of either $\eta$ or $\epsilon$ for the given  ranges of $\bigtriangleup m^2_{21}$ and 
$\bigtriangleup m^2_{23}$ which are consistent with observational data[1].

For a demonstration, we present here the   numerical estimations for the value of 
solar mixing $\tan^2\theta_{12}=0.45$
which in turn corresponds to two values of $r=\eta/\epsilon$  at $r=0.840498455$ and $r=3.1594955$ 
derived  from eq.(9). The first value leads to   two ranges of $\eta$ as 
(A): $0.00398619\leq \eta\leq 0.00527107$ and (B):$0.586525 \leq\eta\leq 0.58781$,
 whereas second one has only one range 
of $\eta$ as (C): $-0.0193327 \leq\eta\leq -0.0147069$. As discussed before, case (B) 
belongs to inverted hierarchy
(Type-IHB) and cases (A,C) belong to Type-IHA. However the expressions for eigenvalues and solar 
mixing angle are the same. We use standard procedure to estimate neutrino masses and mixings[10,11].

{\bf Case (A): Type-IHA where mass eigenvalues are of the form $(m_1, m_2, m_3)$.}\\
 Using  the best value $\eta=0.00462863$ and $\epsilon=0.0055071$, we have  
 \begin{equation}
m_{LL}= M_{IHA}=
\left(\begin{array}{ccc}
0.0494493  & -0.000275351 & -0.000275351 \\
-0.000275351 & 0.025  & 0.0247686  \\
-0.000275351 & 0.0247686  & 0.026
\end{array}\right)
\end{equation}
Diagonalising the above mass matrix we have three  mass eigenvalues,
$$ m_i= (0.0491881, 0.0500298, 000231432)eV, \ \ \ i=1, 2, 3.$$
 leading  to $\Delta m^2_{21}=8.35\times 10^{-5}eV^2$ and $\Delta m^2_{23}=2.5\times 10^{-3}eV^2$.
The MNS mixing matrix is extracted as ,
\begin{equation}
U_{MNS}=
\left(\begin{array}{ccc}
0.830455  & 0.557086 & -2.5\times 10^{-18} \\
0.393919 & -0.58722  & -0.707107  \\
0.393919 & -0.58722  & 0.70717
\end{array}\right)
\end{equation}
 which gives $\tan^2\theta_{12}=0.45$, $\tan^2\theta_{23}=1$ and $\sin\theta_{13}=0$.

{\bf Case (B):Type-IHB where the mass eigenvalues are of the form $(-m_1, m_2, m_3)$.}\\
For the best input value $\eta=0.5871675$ and $\epsilon=0.6985945$ we have  
 \begin{equation}
m_{LL}= M_{IHB}=
\left(\begin{array}{ccc}
-0.0198595  & -0.0349297 & -0.0349297 \\
-0.0349297 & 0.025  & -0.00435837  \\
-0.0349297 & -0.00435837  & 0.026
\end{array}\right),
\end{equation}
Diagonalising the above mass matrix we have three  mass eigenvalues,
$$ m_i= (-0.0529967, 0.0537789, 0.0293584)eV, \ \ \ i=1, 2, 3.$$
 leading to $\Delta m^2_{21}=8.35\times 10^{-5}eV^2$ and $\Delta m^2_{23}=2.03\times 10^{-3}eV^2$.
 The model is quite  different from degenerate model where 
the overall magnitude of neutrino masses, is of the order of $0.4$eV.
The MNS mixing matrix is extracted as ,
\begin{equation}
U_{MNS}=
\left(\begin{array}{ccc}
-0.830455  & 0.557086 & -4.4\times 10^{-17} \\
-0.393919 & -0.58722  & -0.707107  \\
-0.393919 & -0.58722  & 0.70717
\end{array}\right),
\end{equation}
 which gives $\tan^2\theta_{12}=0.45$, $\tan^2\theta_{23}=1$ and $\sin\theta_{13}=0$.
 
{\bf Case (C):Type-IHA where the mass eigenvalues are of the form $(m_1, m_2, -m_3)$.}\\
For the best input value  $\eta=-0.0170198$ and $\epsilon=-0.005386865$ we have, 
 \begin{equation}
m_{LL}= M_{IHA}=
\left(\begin{array}{ccc}
0.0505387  & 0.000269343 & 0.000269343 \\
0.000269343 & 0.025  & 0.025851  \\
0.000269343 & 0.025851  & 0.025
\end{array}\right)
\end{equation}
Diagonalising the above mass matrix we have three  mass eigenvalues
$$ m_i= (0.0502832, 0.0511065, -0.00085099)eV, \ \ \ i=1, 2, 3.$$
 leading to $\Delta m^2_{21}=8.3\times 10^{-5}eV^2$ and $\Delta m^2_{23}=2.6\times 10^{-3}eV^2$.
The MNS mixing matrix is extracted as ,
\begin{equation}
U_{MNS}=
\left(\begin{array}{ccc}
0.830455  & -0.557086 & -1.5\times 10^{-18} \\
-0.393919 & -0.58722  & -0.707107  \\
-0.393919 & -0.58722  & 0.70717
\end{array}\right)
\end{equation}
 which gives $\tan^2\theta_{12}=0.45$, $\tan^2\theta_{23}=1$ and $\sin\theta_{13}=0$.

We present our calculations in Table-1 for all possible ranges of $\eta$ and $\epsilon$
leading to different values of solar mixing $\tan^2\theta_{12}$ at $0.5$ for tri-bimaximal 
mixing,  and then  
 $0.45$ and $0.35$ as possible deviations from tri-bimaximal mixing.
The present analysis shows  enormous  provision for  lowering  the solar mixing angle  
without sacrificing predictions on  $\tan^2\theta_{23}=1$ and $\sin\theta_{13}=0$. It is interesting 
to note   that only two parameters $\eta$ and $\epsilon$ play the key roles in the  whole analysis.
For each value of $\tan^2\theta_{12}$ we have three  solutions  corresponding to two values of  $\eta/\epsilon$,
 and every solution has a particular range of $\eta$. These values  satisfy  observed ranges of 
 $\Delta m^2_{21}$ and $\Delta m^2_{23}$.

As a representative example, we present in Fig.1 the graphical solution of the ratio $\eta/\epsilon$ corresponding to 
$\tan^2\theta_{12}=0.45$. In Figs. 2-4 we  summarise all the  results of the calculation corresponding to   
$\tan^2\theta_{12}=0.45$ case.  In  particular, Fig.2  presents a graphical  summary of the predictions on 
$\Delta m^2_{21}$ and   $\Delta m^2_{23}$, and  a corresponding   correlation graph between them for the valid  range
 $0.003986 \leq \eta \leq 0.005271$. Similarly, Figs. 3 and 4 for $0.586525 \leq \eta \leq 0.58781$ and 
 $-0.019333 \leq \eta \leq -0.014707$ ranges, respectively.

\section{Summary and discussions}
We summarise the main points in this work. The inverted hierarchical neutrino mass matrix having 2-3 symmetry,
has great potential to give  tri-bimaximal mixings and also possible deviations of solar mixing 
angle, without sacrificing predictions on maximal atmospheric mixing angle and zero reactor angle.
We do not take  corrections from charged lepton mass matrix and this helps to maintain 2-3 symmetry of the neutrino mass matrix. 
Such  neutrino mass matrices are generated from seesaw formula 
using  diagonal form of Dirac mass matrix and non-diagonal form of right 
handed Majorana mass matrix. Two types of inverted hierarchical models
 are found  
to be crossed over at  a particular range of input parameters. 
The present analysis though phenomenological,  may have important 
implications in 
model buildings[22] on tri-bimaximal mixings 
and its possible deviations,  based on various discrete symmetries  as well as 
non-Abelian gauge groups, and also on the experimental requirements for mass hierarchy 
measurements at zero reactor angle[5].

{\bf Table-1: }
Prediction of the solar mixing angle and its deviation from tri-bimaximal 
mixings,  along with the predictions on solar and atmospheric mass-squared differences.

\begin{tabular}{lllll}\\ \hline
$\tan^2\theta_{12}$ & $\eta/\epsilon$ & range of $\eta$ & 
      $\bigtriangleup m^2_{21}(10^{-5}eV^2)$ & $\bigtriangleup m^2_{23}(10^{-3}eV^2)$ \\ \hline
$0.5$ & $1.0$  &  $0.004835-0.006395$ & $7.20-9.50$ & $2.50-2.50$ \\
$0.5$ & $1.0$ &  $0.66072-0.66183$      & $9.50-7.20$ & $1.41-1.41$ \\ 
$0.5$ & $3.0$  &   $-0.01871 - -0.01423$ & $ 9.41-7.20$ & $2.63-2.60$ \\ \hline
$0.45$ & $0.84049$  &  $0.003986-0.005271$ & $7.17-9.50$ & $2.5-2.5$ \\
$0.45$ & $0.84049$ &  $0.586525-0.58781$      & $9.5-7.2$ & $2.03-2.03$ \\ 
$0.45$ & $3.15949$  &   $-0.019333 - -0.014707$ & $ 9.41-7.20$ & $2.63-2.60$ \\ \hline
$0.35$ & $0.44620$  &  $0.002002-0.0026463$ & $7.20-9.50$ & $2.51-2.51$ \\
$0.35$ & $0.44620$ &  $0.36217-0.362811$      & $9.5-7.2$ & $4.01-4.01$ \\ 
$0.35$ & $3.55380$  &   $-0.020592 - -0.015666$ & $ 9.41-7.20$ & $2.63-2.60$ \\ \hline
\end{tabular}

\section*{Acknowledgements}
One of the authors(NNS) is thankful to organisers of the IX Workshop on High Energy Physics 
Phenomenology (WHEPP-9) held at Institute of physics, Bhubaneswar, India, January 3-14, 2006, 
where the present work  was initiated. Useful interactions with Prof.G. Altarelli and  Prof.Ernest Ma 
during the workshop are highly acknowledged.

\pagebreak

\begin{figure}[!h]
\begin{center}
\includegraphics[width=15cm]{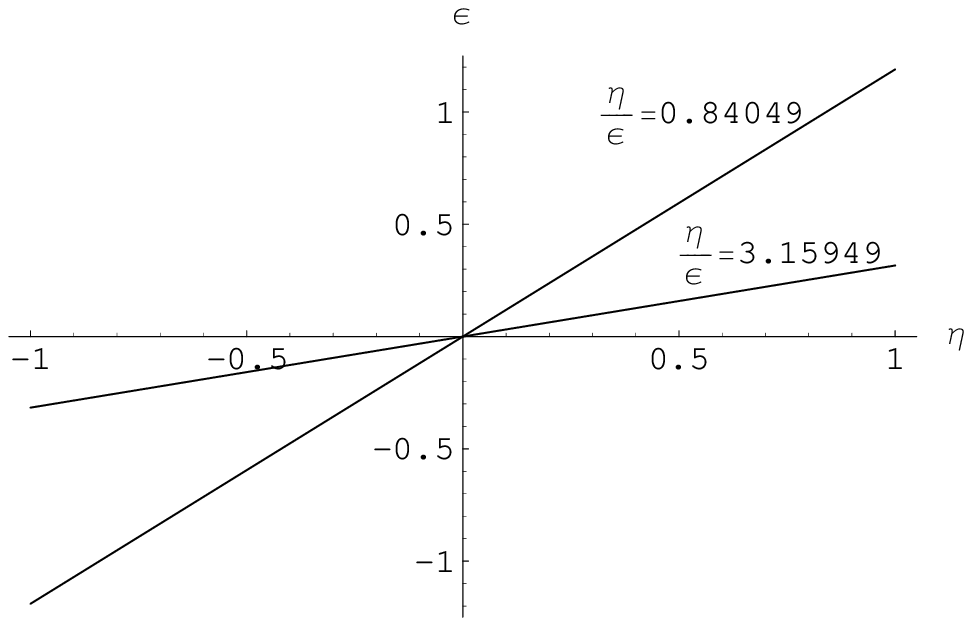}
\caption{Two solutions of $\tan^{2}\theta_{12}=0.45$ corresponding to $(\eta / \epsilon)$ equals to $0.84049$ and $3.15949$ respectively.}
\end{center}
\end{figure}

\makeatother
\begin{figure}
\begin{center}\includegraphics[%
            width=0.50\textwidth]{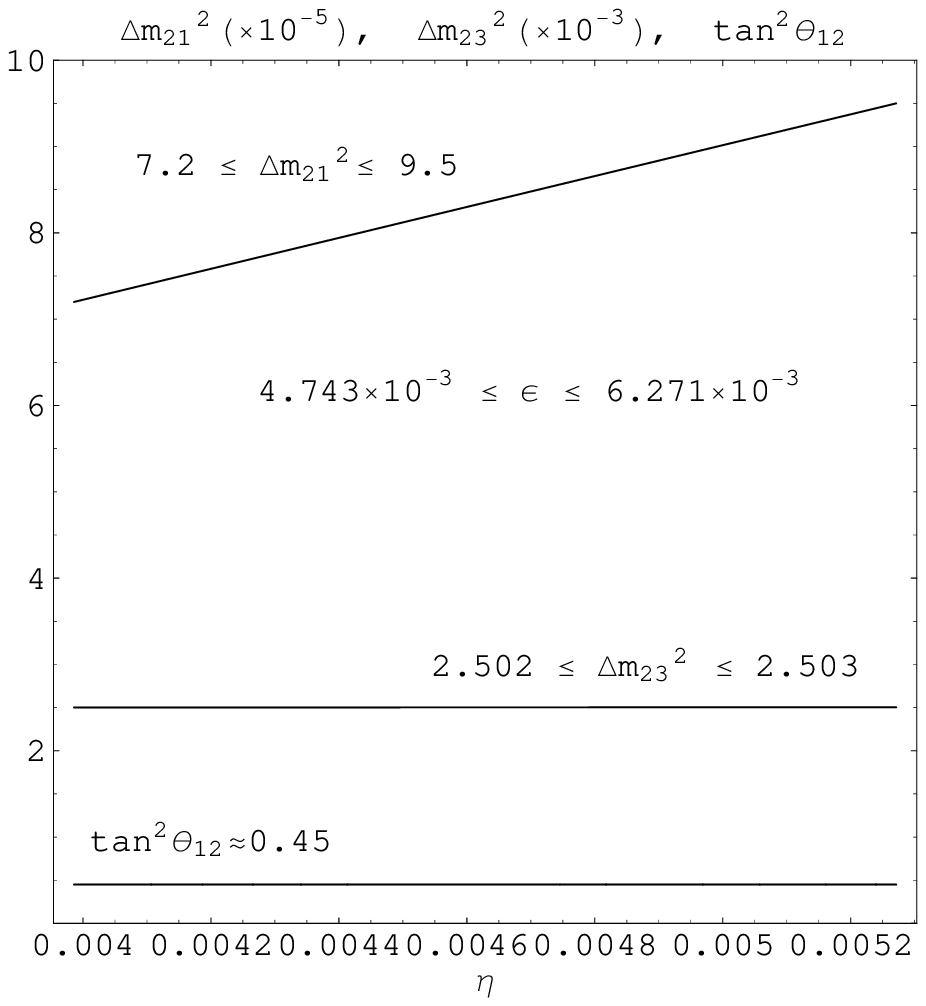}\hfill{}\includegraphics[%
            width=0.50\textwidth]{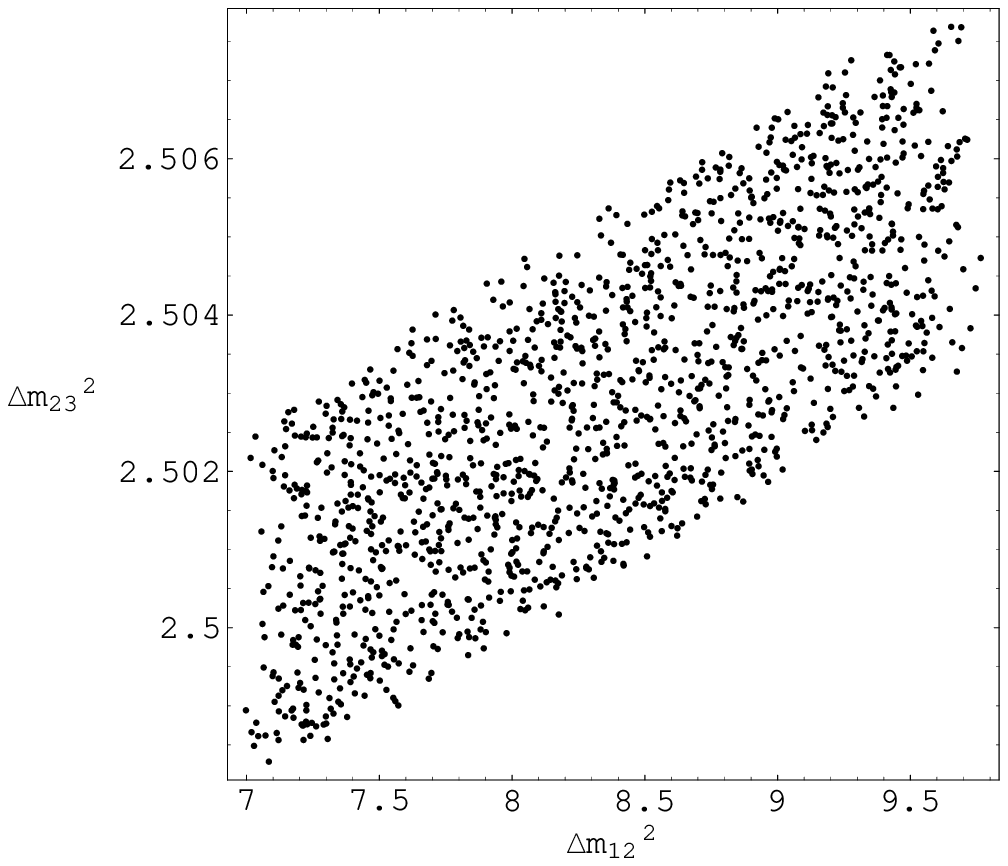}
\end{center}
\caption{Predictions on  $\Delta m^{2}_{21}$ in the unit $(10^{-5}eV^2)$ and 
 $\Delta m^{2}_{23}$ in the unit  $(10^{-3}eV^2)$ for the value  $\tan^2\theta_{12}=0.45$ in the  range:
$0.003986 \leq \eta \leq 0.005271$
 and the corresponding correlation graph.}
\end{figure}
\begin{figure}
\begin{center}\includegraphics[%
            width=0.50\textwidth]{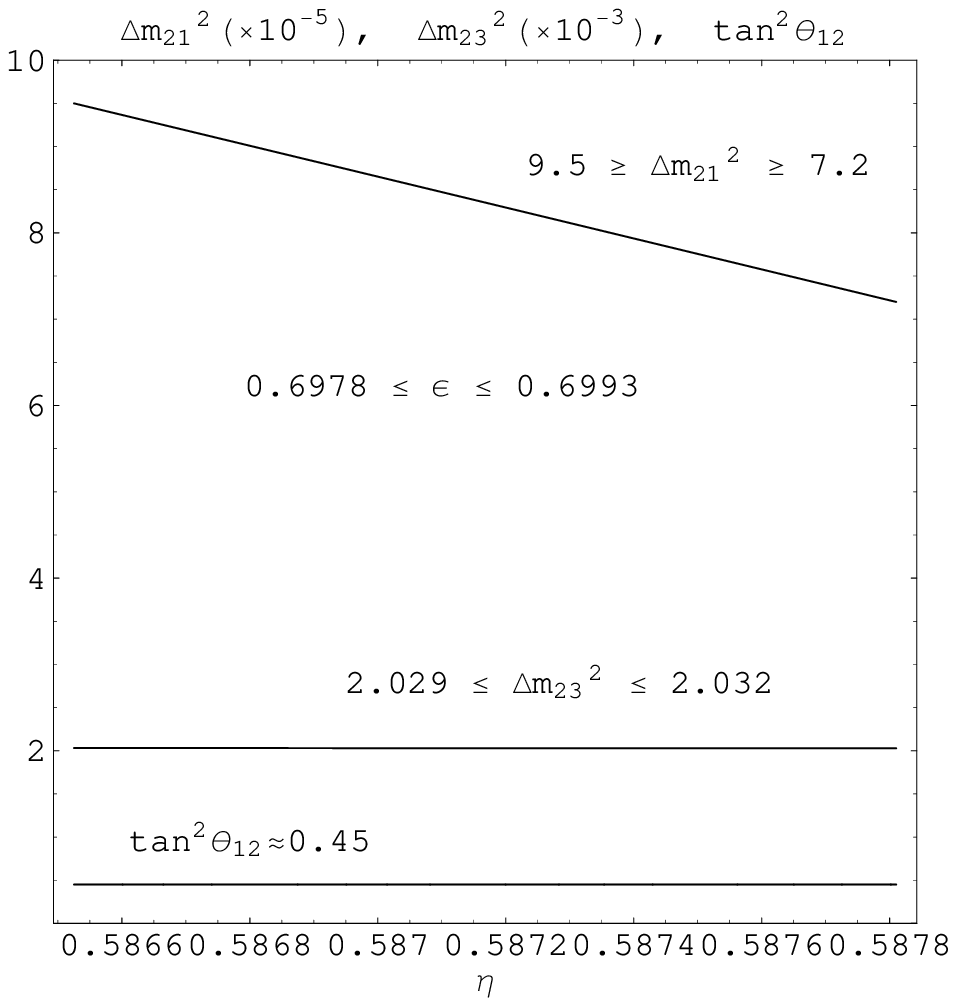}\hfill{}\includegraphics[%
            width=0.50\textwidth]{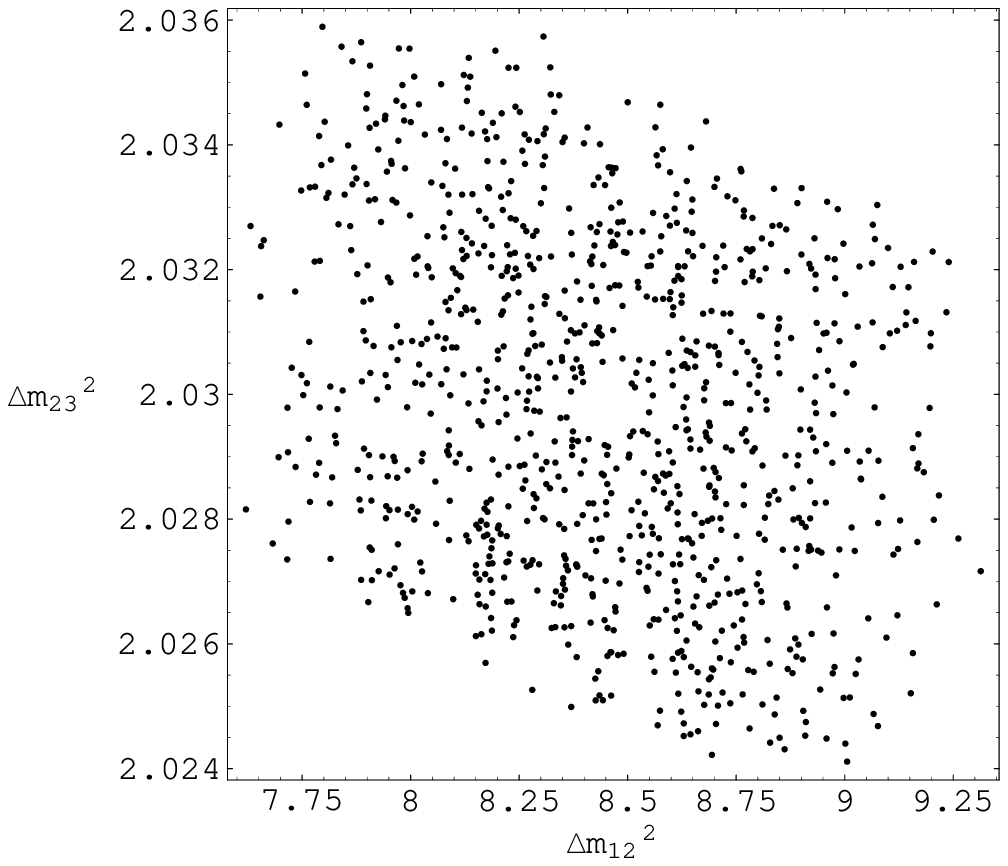}
\end{center}
\caption{Predictions on  $\Delta m^{2}_{21}$ in the unit $(10^{-5}eV^2)$   and 
 $\Delta m^{2}_{23}$ in the unit $(10^{-3}eV^2)$    for the value  $\tan^2\theta_{12}=0.45$ in the valid range:
 $0.586525 \leq \eta \leq 0.58781$ and the corresponding correlation graph.}
\end{figure}
\begin{figure}
\begin{center}\includegraphics[%
            width=0.50\textwidth]{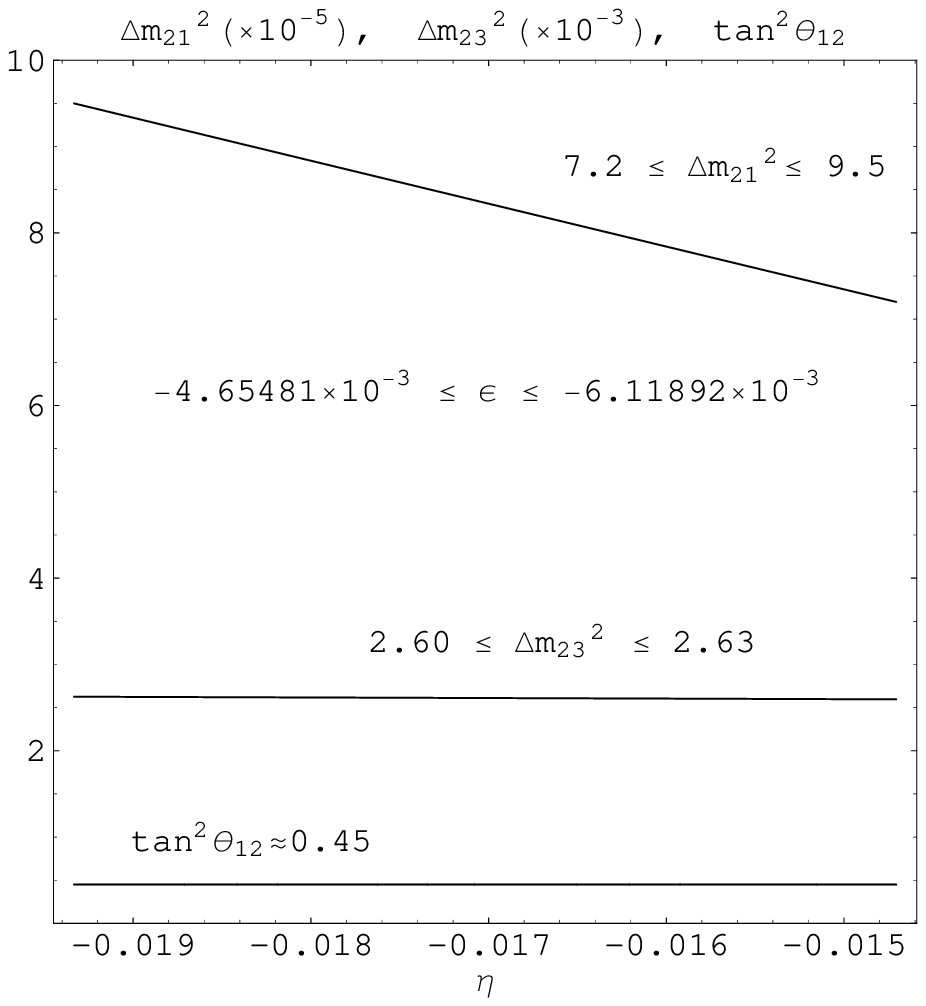}\hfill{}\includegraphics[%
            width=0.50\textwidth]{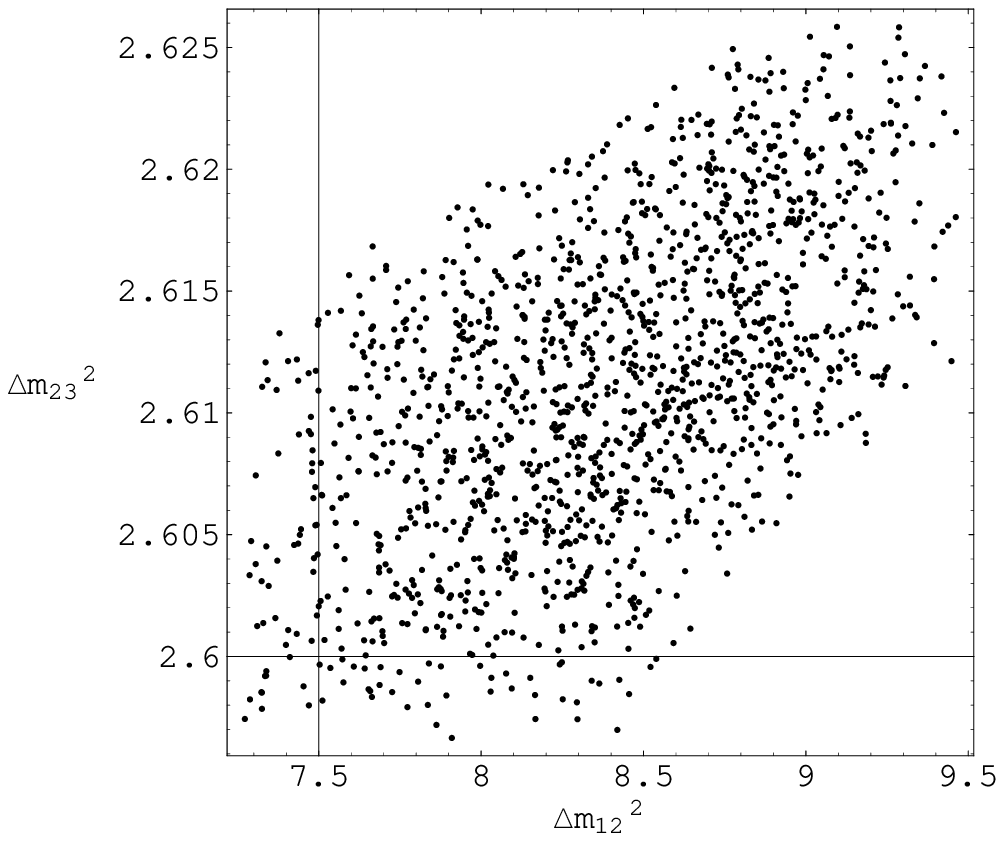}
\end{center}
 \caption{Predictions on  $\Delta m^{2}_{21}$ in the unit $(10^{-5}eV^2)$  and 
 $\Delta m^{2}_{23}$ in the unit $(10^{-3}eV^2)$   for the value  $\tan^2\theta_{12}=0.45$ in the valid range: 
$-0.019333 \leq \eta \leq -0.014707$ and the corresponding correlation graph.}
\end{figure}


\end{document}